\documentclass{article}
\input epsf.sty
\newcommand{\bright}{\begin{flushright}}
\newcommand{\eright}{\end{flushright}}
\newcommand{\bminip}{\begin{minipage}}
\newcommand{\eminip}{\end{minipage}}
\newcommand{\bcent}{\begin{center}}
\newcommand{\ecent}{\end{center}}

\newcommand{\Lmd}{\Lambda}

\newcommand{\lsim}{\mbox{\raisebox{-.3em}{$\;\stackrel{<}{\sim}\;$}}}

\newcommand{\beq}{\begin{equation}}
\newcommand{\eeq}{\end{equation}}
\newcommand{\bfig}{\begin{figure}}
\newcommand{\efig}{\end{figure}}

\begin{document}
\baselineskip=0.6cm
\bcent
{\Large\bf A possible new interpretation of the results of $\Delta\alpha/\alpha$ from QSO absorption lines} \\[.6em]
Yasunori Fujii\\[.0em]
{\small Advanced Research Institute for Science and Engineering,Waseda University, Tokyo, 169-8555 Japan}
\mbox{}\\[.8em]
{\bf Abstract}
\ecent
\hspace*{1cm}
\bminip{13.5cm}
\baselineskip=0.5cm
The measurement of $\Delta\alpha/\alpha$ from the QSO absorption lines of Fe {\footnotesize II} taken at two different redshifts (Quast et al, Levshakov et al) is compared with the result based on the 23 data points also obtained from VLT/UVES (Chand et al), to demonstrate that the apparent oscillatory time-dependence fitting the latter taken at face value is reinforced by the former claimed to be of much better accuracy.
\eminip
\baselineskip=0.6cm
\mbox{}\\[.6em]

Results of the two groups for the possible time-variability of the fine-structure constant derived from the QSO absorption lines appear to be incompatible with each other.  The Keck/HIRES group  reports a non-zero result, $\Delta\alpha/\alpha = (-0.573\pm 0.113)\times 10^{-5} $(MFWDPW \cite{143}), whereas one of the VLT/UVES measurements produces an apparently null result, $ (-0.06\pm 0.06)\times 10^{-5}$ for the same quantity (case 1 of CSPA \cite{23}).  Another VLT/UVES group shows $(0.04\pm 0.15)\times 10^{-5}$ (QRL \cite{QRL} and LCMD \cite{LCMD}), thus favoring case 1 of CSPA (to be called CSPA1 in what follows).  All of these groups relied on the many-multiplet method  and analyzed the results in terms of the weighted-mean estimate, which is interpreted as a 1-parameter fit in terms of a constant $\Delta\alpha/\alpha$.

We proposed, in contrast, an oscillatory behavior for $\Delta\alpha/\alpha$ as a function of time based on the scalar field supposed to be responsible for the acceleration of the universe \cite{cup}; the required ``tracking" behavior can be realized by the scalar field being trapped to a potential minimum, thus causing a damped oscillation.  The oscillation is supported also by the Oklo constraint indicating a much smaller $\Delta\alpha/\alpha$ at the look-back time $s\approx  0.142$ \cite{shl,YFET}, somewhat closer to the present time than the range of the QSO observation, $0.3 \lsim s \lsim 0.8$.  The meteorite constraint at $s\approx 0.34$, however, is not included because of the likely model-dependence \cite{YFAI}.  From a phenomenological point of view, we tried to fit the QSO data in terms of a damped oscillator with 3 parameters.

We applied the 3-parameter fit in this sense to MFWDPW \cite{PL,ptb} and CSPA \cite{mzn}. For the former, we found nearly the same degree of fit as in the original weighted-mean analysis, whereas we came across a nonzero fit to CSPA1, even with a considerably ``improved" $\chi_{\rm rd}^2$, though its value 
much smaller than 1 might indicate a poor statistical quality of the data used in this particular manner.

It can be argued \cite{LCMD}, on the other hand, that the real accuracy in the analysis in CSPA may not have reached the level as presented in \cite{23}.  It may not be sufficient to detect detailed structure of the behavior, if any, of $\Delta\alpha/\alpha$.  We pointed out \cite{mzn}, on the other hand, that choosing the 1- or 3-parameter fit depends on what type of distribution, flat or oscillatory, we expect to be approximated from a theoretical point of view.  An oscillatory behavior might be easily averaged out by the assumed flat 1-parameter fit, for example.  In Ref. \cite{mzn}, we in fact demonstrated that the CSPA1 data allows a nonzero oscillatory fit not in contradiction with a null result which could be derived by a 1-parameter fit.  In this sense, the oscillatory fit may deserve further scrutiny for its own, not being limited strictly to the parameters derived currently from CSPA1.

In this article we apply the same 3-parameter approach to QRL and LCMD, finding that these results, claimed to have been analyzed much more carefully than the others, appear to provide a strong support for the oscillatory nature of CSPA1.  It even seems as if the accuracy of the measurement and analysis has been already sufficient to describe the physical content correctly.  This might be too optimistic a view from current efforts to remove many kinds of systematic uncertainties by using larger telescopes, for example, not to mention a very limited number of data points utilized to draw our conclusion.  We still emphasize that our numerical result obtained from the available data appears too impressive to be ignored as a mere coincidence, as long as we take the CSPA1 result at face value.

We start with our 3-parameter fit of a damped-oscillator to $y(s)=(\Delta\alpha/\alpha) \times 10^5$ as a function of the look-back time $s=1-t/t_0$ with $t_0$ for the present age of the universe \cite{ptb}:
\beq
y(s)= a\left(  e^{b(s-1)} \cos(v-v_1)-e^{-b} \cos(v_1) \right),
\label{lev-1}
\eeq
where $v/s = v_1/s_1=v_{\rm oklo}/s_{\rm oklo}=2\pi T^{-1}$ with the coefficient $v_1$ determined by
\beq
v_1 = \tan^{-1}\left( \left( e^{-bs_{\rm oklo}} -\cos(v_{\rm oklo})\right) /\sin(v_{\rm oklo}) \right),
\label{lev-2}
\eeq
due to which $y(s)$ vanishes at $s=0$ (today) and $s=s_{\rm oklo}\approx 0.142$ corresponding to the Oklo time about $1.95\times 10^9 {\rm Gy}$ ago.  We modified the old definition $a_{\rm old}$ of the amplitude in \cite{PL,ptb,mzn} by $a= a_{\rm old} e^b$ for an easier representation of Fig. 2 in the following.

\bfig[htb]
\epsfxsize= 10cm
\hspace*{2cm}
\epsffile{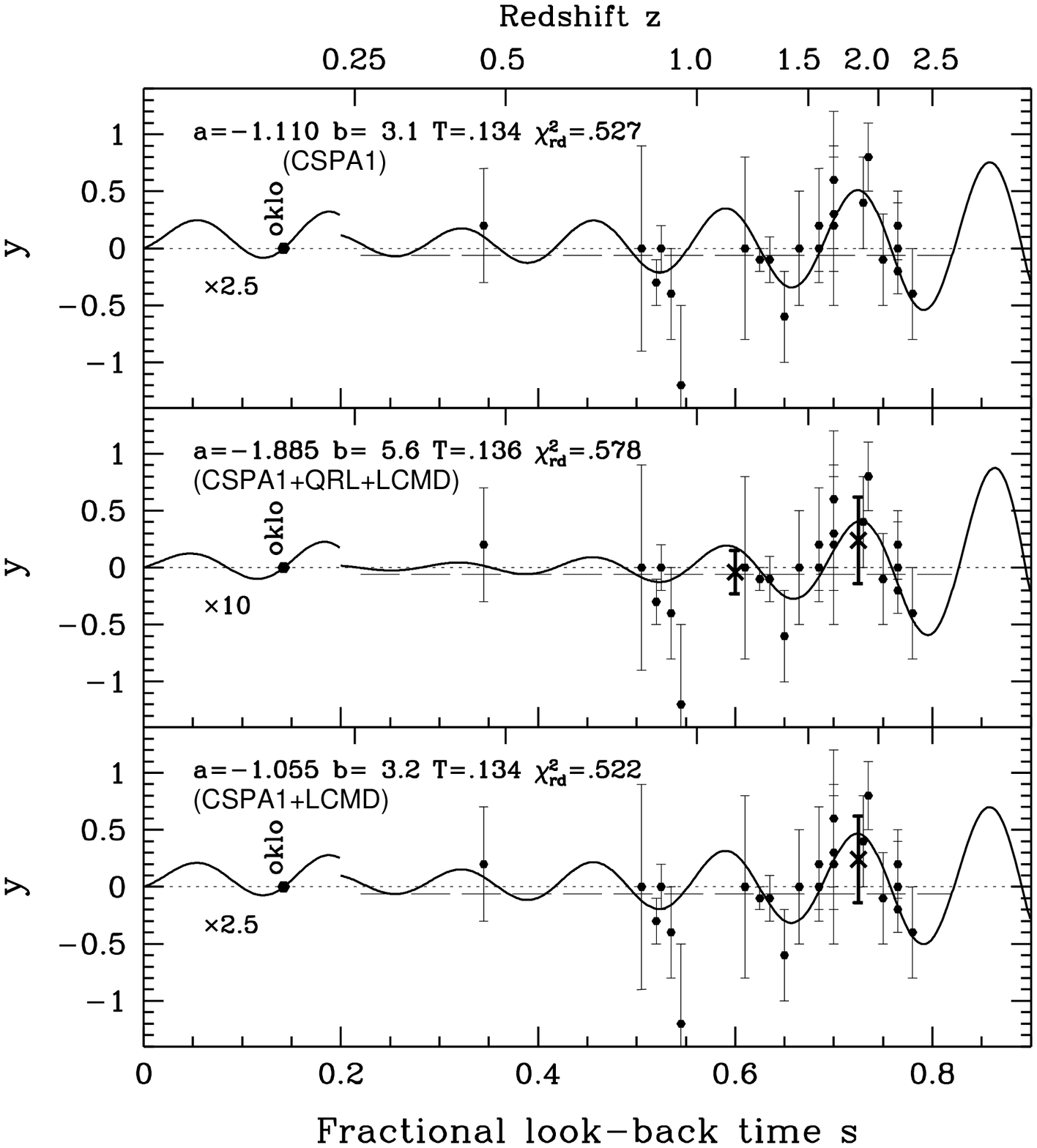}
\caption{The best 3-parameter fits for CSPA1 only (upper panel), CSPA1 + QRL + LCMD (middle panel), and CSPA1 + LCMD (lower panel), versus the fractional look-back time $s$ as well as the redshift $z$.  The curves below $s=0.2$ are magnified as indicated in each panel, including the (assumed) zero at $s_{\rm oklo}=0.142$.  The points for QRL, LCMD are marked by crosses with the error bars denoted by thicker lines.  The horizontal dashed line stands for $-0.06$ for the weighted mean derived in CSPA1.}
\label{VLTLV1_1}
\efig
\bfig[htb]
\epsfxsize= 10cm
\hspace*{2cm}
\epsffile{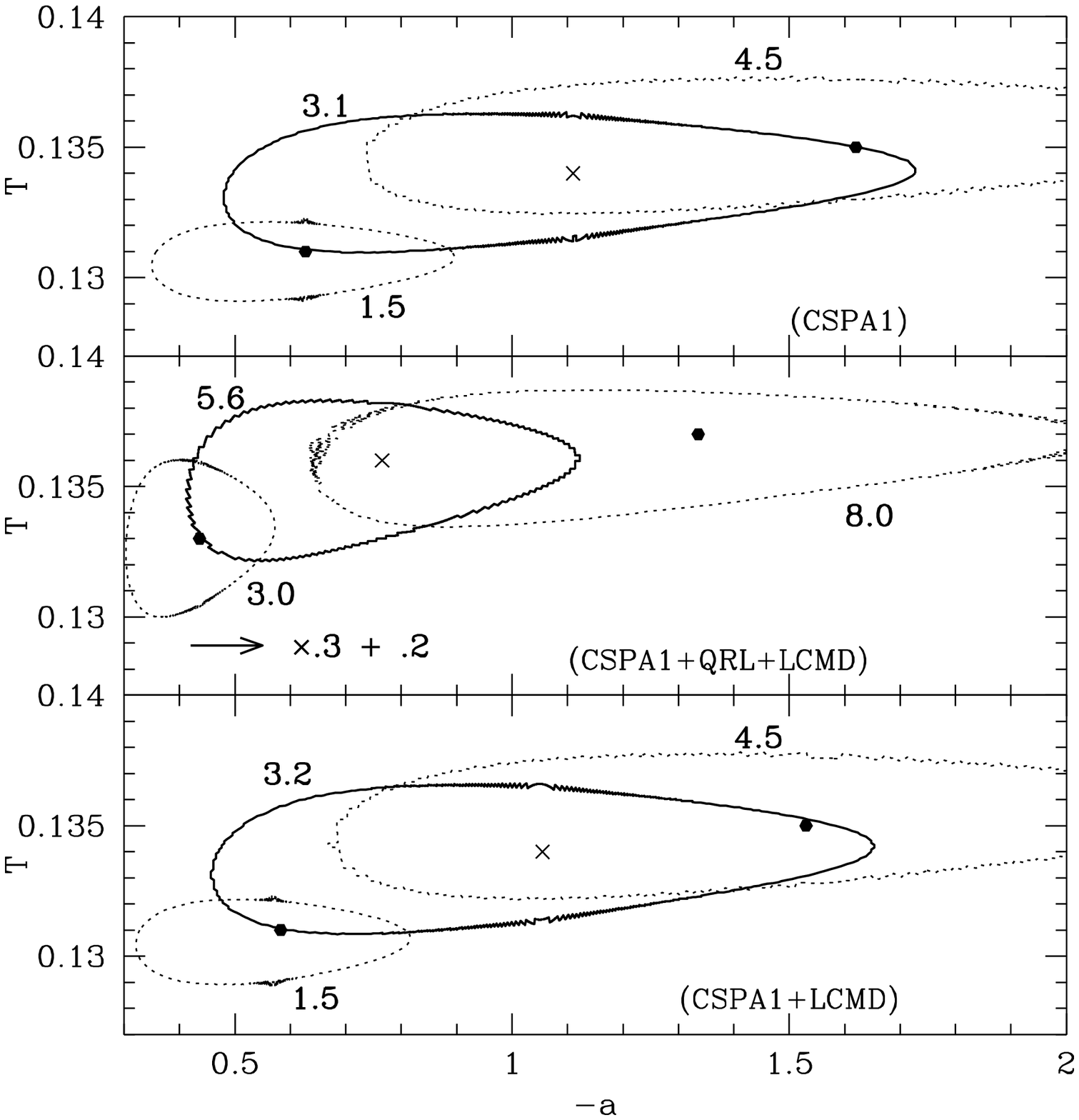}
\caption{The 68\% confidential volumes in 3-dimensional $a,b,T$ space are shown as 2-dimensional cross sections for each value of $b$, as indicated alongside the corresponding contours.  The contours drawn by solid curves also with the crosses inside correspond to the best values of $b$, whereas those by dotted curves with filled circles are for representatives of different values of $b$.  The three panels correspond to those in Fig. 1, respectively.  In the middle panel, the horizontal scale has been stretched by the ratio 0.3 and shifted by 0.2  to the right.
}
\label{VLTLV1_2}
\efig

The upper panel of Fig. 1 for the CSPA1 result has been borrowed from Fig. 3 of \cite{mzn} with $a=-0.050$ replaced by $a\times e^{3.1}= -1.11$.  The redshift $z$ in the observation has been translated to $s$ by assuming the cosmological expansion parametrized by $t_0= 1.38\times 10^{10}{\rm y}, h=0.7, \Omega_{\Lmd}=0.7$.   In addition to the natural zero at $s=0$, we chose the vanishing at $s=s_{\rm oklo}$ supposed to be a good approximation for much smaller Oklo constraint $|y|\lsim 10^{-2}$ \cite{YFET}, which likely remains robust even with the suggested non-Maxwell-Boltzmann neutron spectrum \cite{lam}.  We found a considerably ``improved" result $\chi_{\rm rd}^2=0.527$ compared with the original $0.95$ for the weighted mean in \cite{23}.

In spite of a critical view against $\chi_{\rm rd}^2 \ll 1$, we notice that the calculated $\chi_{\rm rd}^2$ is small simply because our fitting curve passes nearly through a number of data points particularly in the range $0.60\lsim s \lsim 0.78$ corresponding to the range of the redshift $1.2\lsim z\lsim 2.4$, where the data shows a pronounced ``peak."  It is tempting to accept this peak as representing something real, though its legitimacy depends crucially on the assumed ``sufficient" accuracy reached already, as discussed before.

The authors of \cite{QRL,LCMD} chose Fe {\footnotesize II} intending to overcome uncertainties coming from the isotopic abundances and the radial velocities, among other things.  By careful analyses they obtained the results for the two redshifts 1.15 and 1.84 with the statistical and systematic errors shown separately.  They {\em combined}, however, the two results to obtain the overall weighted mean $y= 0.04 \pm 0.15_{\rm stat}$, which is in favor of CSPA1, but not MFWDPW.  From our point of view of the oscillating behavior, we attempt to consider their two results {\em separately} instead of combining them together.

We first consider the 25 data points consisting of CSPA1, QRL( $y=-0.04\pm 0.19_{\rm stat}$ at $z=1.15, s=0.60$) and LCMD($y=0.24\pm 0.38_{\rm stat}$ at $z=1.84, s= 0.725$).   Our best 3-parameter fit is shown in the middle panel of Fig. 1, together with the corresponding 68\%-confidence region, also in the middle panel of Fig. 2.  The best value of $T$ turns out to be nearly the same as in the upper panel.  The values of $b$, and hence of $a$ are somewhat different.  We also find $\chi_{\rm rd}^2= 0.578$ which is larger than 0.527 for the pure CSPA1 slightly.  This suggests that the Fe {\footnotesize II} data shares some similarity with CSPA1.

We find in fact that the LCMD data point, in particular, lies in the midst of the surrounding points that constitute an obvious ``peak" of CSPA1.  It appears as if the LCMD result obtained with much more care than CSPA1 reinforced the most important part of the oscillating nature as was observed in CSPA1.  In order to  support this view further, we applied the same 3-parameter fit to the 24 data points including the CSPA1 and LCMD data, as shown in the lower panels of Figs. 1 and 2.  The similarity with the pure CSPA1 is even more obvious, resulting in the still smaller value $\chi_{\rm rd}^2 = 0.522$.  This implies, on the other hand, that the QRL result contributes a little to make the agreement poorer.  We still point out that the fitting curve in the middle panel of Fig. 1 passes barely off the 1 standard deviation at $z=1.15$ (QRL), implying the ``damage" to be by no means serious.

We also add that the result of CSPA1 together with QRL alone can be fitted by $a= -2.18, b= 5.9, T= 0.136, \chi_{\rm rd}^2= 0.595$.  As anticipated, the reduced chi-squared is a little larger than that corresponding to the middle panel of Fig. 1, but not significantly again.

To summarize, we emphasize that the Fe {\footnotesize II} samples due to QRL and LCMD, taken at face value at the separate $z$ values, reinforce the oscillatory behavior as illustrated in CSPA1.  We admit, however, that the strongest support comes from a {\em single} data point at $z=1.84$ (LCMD).  We still notice
 that this data point happens to occur at $z$ around which we find the peak behavior crucially important to reveal the oscillatory structure in CSPA1.  This latter analysis might be subject, however, to future improvement with respect to the accuracy, as was argued in the most recent version (v3) of LCMD.  We must be open to a more conventional wisdom that the required accuracy has not been fully realized yet to establish the presence of a peak unambiguously.  In order to determine if the oscillatory time-dependence of $\alpha$ is in fact justified, we urge that an emphasis should be placed in the future analysis to make careful measurement at {\em different} $z$ values even for relatively sparse data points.

We close the paper by adding a theoretical remark.  Once we accept an oscillating behavior, it is irrelevant to ask if $\alpha$ was larger or smaller in the past.  As a consequence, rather than dividing the past flat value of $y$ obtained from the weighted-mean by an averaged time span, we estimate today's rate of change  of $\alpha$ in terms of the tangential slope of $y(s)$ at $s=0$; $(\dot{\alpha}/\alpha)_0 =-t_0^{-1}y'(0)\times 10^{-5}$ \cite{mzn}.  The upper and the lower panels in Fig. 1 yield a particularly ``large" value, $-0.96 \times 10^{-15}{\rm y}^{-1}$, which is close to the upper bound $\sim 2\times 10^{-15}{\rm y}^{-1}$ reached recently by the laboratory experiment \cite{peik}.

I would like to thank Paolo Molaro for his valuable remarks on the nature of the data of QRL and LCMD.  I also thank Aldo Fiorenzano and Naoto Kobayashi for their illuminating discussions on the subject.

\end{document}